\documentclass[]{article}
\newtheorem{theorem}{Lemma}
\begin{document}
\title{Nondimensional Simplification of Tensor Polynomials with Indices}
\author{
 A. Balfag\' on\thanks{Institut Qu\'\i mic de Sarri\` a, Laboratori de F\'\i sica Matem\` atica,
Societat Catalana de F\'\i sica (I.E.C.)Universitat Ram\' on Llull e-mail: abalf@iqs.url.es} and X. Ja\' en.\thanks{Universitat Polit\` ecnica de Catalunya, Laboratori de F\'\i sica Matem\` atica,
Societat Catalana de F\'\i sica (I.E.C.)
e-mail: jaen@baldufa.upc.es}
}
\date{}
\maketitle
\begin{abstract}
We are presenting an algorithm capable of simplifying tensor polynomials with indices when the building tensors have index symmetry properties. These properties include simple symmetry, cyclicity and those due to the presence of covariant derivatives.
The algorithm is part of a {\it Mathematica} package called {\it Tools of Tensor Calculus (TTC)}[web address: http://baldufa.upc.es/ttc]
\end{abstract}
\section{Introduction}
The two main languages commonly used in writing tensor calculus expressions are the intrinsic notation and the index notation. The intrinsic notation seems to be the preferred language for  making computer algorithms  work with symbolic tensor expressions.  The index notation, on the other hand, is extremely powerful in expressing and manipulating tensors. There are some simple expressions (in index notation) which are difficult to express in intrinsic notation for the purpose of introducing new operators with new properties. This is the case e.g. expressing
the simple operation of raising an index  (\ref{eq1})
\begin{equation}
T_{i\ k}^{\ j}=g^{j m}T_{i m k} \label{eq1}
\end{equation}
Recently  some computer tools capable of  working with index notation at a symbolic level have appeared \cite{MathTensor,Ilyin,Balfagon}.

In the present study we are facing the problem of tensor polynomial simplification using index notation.
In a recent article \cite{Balfagon} we dealt with this problem, restricted to the case of simple index symmetry properties of the building tensors like 
\begin{equation}
T_{ i_{\sigma (1)}... i_{\sigma (n)}}+a\,\, T_{i_{1} ...i_{n}}=0\label{mono}
\end{equation}
 $a$ being a scalar and $\sigma$ a permutation of $\{ 1,...,n\}$.
 (\ref{mono}) includes symmetry and antisymmetry of any sequence of indices, and pairsymmetry. 
We will refer to these kinds of properties as {\it monoterm}. The algorithm proposed in \cite{Balfagon} does not use any kind of library, either temporal or permanent, and for that reason it does not require a lot of computer memory. Nevertheless, beside the monoterm restriction the algorithm presented did not draw benefit from all the calculus done. It was reasonably fast on one monomial but not on a hundred, which is a problem to  be solved when using other properties. 

V.A. Ilyn and A.P.Kryukov \cite{Ilyin} have proposed a method capable of working with index symmetry properties, which include cyclic type properties like

\begin{equation}
T_{i j k}+T_{k i j}+T_{j k i}=0\label{cyclic}
\end{equation}
(\ref{cyclic}) was included by  the authors in a set called {\it multiterm} like properties.
Although the problem was theoretically satisfactorily solved, the computer time and the amount of computer memory needed to make elementary simplifications was large enough to consider that this study may have some practical and fundamental limitations ( since an expression with 11 indices needs about 2500Mb). 

One of the problems encountered by V.A. Ilyn and A.P.Kryukov is related to the dummy indices. They consider the possibility of renaming dummy indices as new relations to add to  the set of relations generated using the true index properties of building tensors. So in the monomial $S_{j k} T^{k j i}$ as well as the properties due to the fact that $S$ is a symmetric tensor they add the ones deduced from the fact that the indices $j, k$ are dummy, i.e.,
 \begin{equation}
S_{j k} T^{k j i}-S_{k j} T^{j k i}=0
\end{equation}
The number of this kind of relations increases factorially with the number of dummy indices, so the algorithm needs to work very hard for monomials of interest.
Other authors \cite{Portugal,Fulling} have studied simplification algorithms but these need to be implemented. R.Portugal \cite{Portugal} handles cyclic like properties similar to \cite{Balfagon}. We think that his algorithm can work very well for monoterm properties ( like \cite{Balfagon} does) but not for multiterm ones. The trouble is that although apparently he can decide if a multiterm rule takes advantage using them increasing the lexicographic order of the working monomial, it can happen that this decision must be delayed after analyzing the effect of some chain of multiterm rules applied to the monomial. In general {\it the whole set of monoterm plus multiterm monomial equations must be considered}.  

The algorithm that we present here, written in {\it Mathematica} \cite{Mathematica} language, is a part of the package {\it Tools of Tensor Calculus (TTC)} \cite{Castellvi1,Castellvi2,Balfagon,TTC}( from now on we will call both, the algorithm and the package, {\it TTC}). {\it TTC} can work with monomials whose building tensors have both  monoterm and multiterm properties. In addition to dealing with  dummy indices it also handles  symmetry  (including  the partial derivatives, although it is not recommended because of their non-tensor character and the increase of the corresponding computer time. In this paper we will leave aside partial derivatives ) and antisymmetry of any sequence of indices, pairsymmetry, cyclic type properties and the Ricci commuting relations due to the presence of two or more covariant derivatives.

Taking the Riemann tensor as a paradigm  {\it TTC} can handle monomials that have this tensor as a factor, among others, with the known properties:

\begin{eqnarray}
R_{i j k l}=-R_{i j l k}\label{riemp01} \\
R_{i j k l}=-R_{j i k l}\label{riemp02} \\
R_{i j k l}=R_{k l i j}\label{riemp1} \\
R_{i j k l}+R_{i l j k}+R_{i k l j}=0\label{riemp2}\\
R_{i j k l ;m}+R_{i j m k ;l}+R_{i j l m ;k}=0\label{riemp3}\\
R_{i j k l} \underbrace{_{;m ....;n}}_{k}+a\, R_{i j k l ;}\underbrace{_{n ....;m}}_{k}+({\rm Riemann\  terms})\underbrace{_{;...}}_{k-2}=0\label{riemp4}
\end{eqnarray}
where (\ref{riemp4}) refers to Ricci commuting relations.
We have left  the inclusion of the properties derived from the dimension of the base space, for the Riemann tensor and for the generic case, for a future work \cite{XADim}.

We have applied the algorithm presented in a real calculus ( superenergy tensor problems) finding new relevant results \cite{XACollinson}.

The plan of the present paper is as follows: in section 2 we present the problem and explain our solution from the theoretical point of view. In section 3 we describe how the algorithm deals with Ricci commuting relations. Section 4 is devoted to describe the main function of the algorithm implementation. In the Appendix A we show some examples of how {\it TTC} works in practice.

\section{The simplification problem }
The problem can be stated as follows:

Given a tensor monomial written in index notation together with all index properties of the building tensors, to find if it is zero and, if not, a unique canonical equivalent polynomial. 

The canonical equivalent polynomial is derived from the complete set of monomial relations generated from the exhaustive application of all index properties of the tensor factors over the original monomial. The solving procedure is related to some ordering criterion of the set of monomials.  The concrete ordering criterion is not important, we will assume that we have a {\it unique} ordering criterion.

\subsection*{Free and dummy indices}

Let  $M(I_{F}^{\beta},J_{D}^{\alpha})$ be a given tensor monomial, with $F$ free indices $I_{F}^{\beta}$ $=\{ j_{\beta (1)},...,j_{\beta (F)}\} $ and $D$ dummy indices $J_{D}^{\alpha}$ $=\{ j_{\alpha(1)},...,j_{\alpha(D)}\} $ ,  $\alpha$ and $\beta$ being any permutation of  $\{1,...,D\}$ , $\{1,...,F\}$ respectively. Example:

\[
M(\{ i,k,l\} ,\{ j,m,p\} )=S_{i j k} T^{j \  m}_{\ p} T^{p}_{\ l m}
\]

Due to the meaning of the dummy indices the following identities hold

\begin{equation}
M(I_{F}^{\beta},J_{D}^{\alpha})= M(I_{F}^{\beta},J_{D})\ \ \ \forall \alpha
\end{equation}
$ J_{D}$ being  the canonical ordering  of dummy indices in the monomial. 
This means that there will be many equivalent expressions for a given monomial. This is the first problem to be solved. 

In \cite{Balfagon} we looked at a method in which a monomial could be rewritten in a canonical form with respect to the dummy indices independently of the number of dummy indices  present in the monomial, without using much computer memory. 

Essentially the method consists of using an ordering criterion that can be applied to a monomial without prior consideration of all the monomials to be ordered. The method presented here is an improved version of \cite{Balfagon}, in order to optimize the amount of storing data and computer time, following the criteria:

{\bf 1)}Free indices are renamed: \verb" IndexF[1], IndexF[2],..." all of them in upper position ( contravariant)

{\bf 2)} Dummy indices are renamed \verb" IndexD[1], IndexD[2],..." all of them in upper position ( contravariant)  

{\bf 3)} We find the permutations of free, $\beta$, and dummy, $\alpha$, indices  in order to find the most ordered monomial expression. Note that at this stage the result is  independent of the original order of free indices. We call this {\it globalcodification}:
\begin{equation}
M(I_{F}^{\beta},J_{D}^{\alpha})
\buildrel{globalcodification}\over{\Longrightarrow} M(I_{F},J_{D})=M(I_{F})
\label{globalcodification}\end{equation}

being $I_{F}$ the canonical ordering of  free indices in the monomial.

$ M(I_{F})$ is the skeleton of the monomial. $ M(I_{F})$ carries all the necessary and sufficient information on the monomial independent of the name of dummy and free indices. Different monomial skeletons will be denoted by $ M_1(I_{F}) M_2(I_{F})...$ 

Finally we take the permutation $\beta$ performed on free indices over the {\it globalcodification} of the monomial. The final result is called {\it codification} of the original monomial and the process will we called $(0)$-rules. 
\begin{equation}
M(I_{F}^{\beta},J_{D}^{\alpha})
\buildrel{(0)-{\rm rules}}\over{\Longrightarrow} M(I_{F}^{\beta},J_{D})= M^{(0)}(I_{F}^{\beta})\label{codification}
\end{equation}

Note that the canonical sets $ I_{F}$ and  $J_{D}$ are defined through (\ref{globalcodification}) since they depend on the skeleton of the monomial. Also note that any permutation of free indices of a {\it globalcodification} corresponds to a {\it codification} version of the corresponding permutation of free indices in the original monomial, so when a monomial has many free indices this procedure optimizes computer time and memory.

On the other hand, as we will see later, all properties found during a calculation are stored in {\it codification} version but all monomials with the same  {\it globalcodification} version can use, due to what we have stated on the precedent paragraph,  the corresponding $\beta$-like permutation  version of the stored properties, so the work of finding properties is essentially independent of the order of free indices.

In the following sections we will handle monomials in the form $M^{(0)}(I_{F}^{\beta})$. In this way we will not mention the dummy indices problem anymore.

\subsection*{Tensor properties}

Of all tensor properties we will emphasize those that will be of importance within the algorithm. 
\begin{itemize}
\item {\bf Monoterm properties}: the result of their application over a given monomial is also a monomial. 

\item {\bf Multiterm properties}: the result of their application over a given monomial is a polynomial (more than one term). 

\item {\bf Homogeneous properties}: they are properties of only one tensor, say $T_{ i_1... i_n}$, of the form 
\begin{equation}
T_{ i_1... i_n}+\sum_{j=1}^{M} a_{\sigma_{j}} T_{ i_{\sigma_{j} (1)}... i_{\sigma_{j} (n)}} =0\label{homogeneous}
\end{equation}
$\sigma_{j}$ being  permutations of indices and $ a_{\sigma_{j}}$ scalars, 
in such a way that (\ref{homogeneous}) is invariant under any of the $\sigma_k^{-1}$, that is
\begin{eqnarray}
T_{ i_{\sigma_{k}^{-1} (1)}... i_{\sigma_{k}^{-1} (n)}}+\sum_{j=1}^{M} a_{\sigma_{j}} T_{ i_{(\sigma_{k}^{-1}\circ\sigma_{j}) (1)}... i_{(\sigma_{k}^{-1}\circ \sigma_{j}) (n)}} =\nonumber\\
A_k \left( T_{ i_1... i_n}+\sum_{j=1}^{M} a_{\sigma_{j}} T_{ i_{\sigma_{j} (1)}... i_{\sigma_{j} (n)}}\right)\ \ \forall k=1,...M
\label{invhomogeneous}
\end{eqnarray}
$A_k$  being some scalars. Example: see (\ref{cyclic}).
\item {\bf Inhomogeneous properties }: These are properties relating one or more tensors whose form is not reducible to (\ref{homogeneous}) fulfilling (\ref{invhomogeneous}).
\end{itemize}

The most important example of inhomogeneous properties is the Ricci commuting relation. The simplest case is when we have a vector field $v$, then
\begin{equation}
v_{i;j;k}  - v_{i;k;j}  - v_m \,R_{\,\,\,\,i\,j\,k}^m=0 \label{ricciid}
\end{equation}
$R$ being the Riemann tensor

The advantage of handling homogeneous properties is that we can pick up a term and use it as a pattern to be applied when finding rules over a given monomial. This procedure, due to the homogeneous character, will be equivalent to use all terms in substitutions on the monomial. This is not true for inhomogeneous properties and special treatment must be considered.

On the other hand some inhomogeneous properties can be applied directly {\it iff} the right hand side inherits all the properties carried from the left hand side. 
There is an important  example when defining the Ricci tensor. $R^{m}_{\ i m j}=R_{i j}$ can be applied directly {\it iff} Ricci inherit from Riemann all the resulting properties, for example 
\begin{equation}
R^{m}_{\ i;m}=\frac{1}{2} R^{m}_{m;i}\label{riccibianchi}
\end{equation}
which is a consequence of applying Ricci definition to the original Bianchi II relation.
If we want to use the definition of the Ricci tensor as a direct relation we must include also (\ref{riccibianchi}) as a direct relation, as well as the deduced properties belonging to the Ricci tensor itself, $R_{ij}- R_{ji}=0$. Although such a procedure can be difficult and unnecessary because we can work exclusively with the Riemann tensor, it is very convenient in order to eliminate superfluous indices.

\subsection*{Induced monomial properties}
Given a set of  monomials $M=\{M_1^{(0)}(I_F^{\beta}),
M_2^{(0)} (I_F^{\gamma}),...\}$ and a set of properties $P=\{p_1,...p_k..\}$ of the building tensors, the set of monomial properties induced by $P$, $P(M)$, is the set of properties which arise applying $P$ over $M$ and over {\it all new monomials generated until no new monomials arise}.

The essential idea when simplifying a given monomial 
$ M^{(0)}(I_F^{\beta})$ is to take the set of properties of the building tensors, $P$, and generate the set of monomial properties $P(\{M^{(0)} (I_F^{\beta})\})$, solving them with respect to the ordering criterion and finally applying the corresponding rules onto the original monomial. The trouble is that the set $P(\{M^{(0)}(I_F^{\beta})\})$ can be too large to be reasonably generated and solved.

It is interesting to try to break the set of properties $P$ into two sets, say $P_1$ and $P_2$. Then generate and solve $P_1(\{M^{(0)} (I_F^{\beta})\})$ applying the resulting rules over $M^{(0)}(I_F^{\beta})$ giving, in general, a polynomial $Pol_{(1)}$  built by the set of monomials $\{...,M_i^{(1)}(I_F^{\beta}),...\}$. Then generate and solve $P_2(\{...,M_i^{(1)} (I_F^{\beta}),...\})$ to give a new set of rules to be applied onto the polynomial $Pol_{(1)}$ giving the final result. 

We will call {\it separable properties} the set $P_1$ of $P$ which can be separated from the rest in the sense of the above paragraph. The separability is a possible property of the given set $P$ which must be demonstrated case by case. Properties which affect different tensors or different sets of indices in  the same tensor are clearly separable, but this is not only possibilitu. It is easy to assess whether a given monoterm property is separable or not. The most important case is the Riemann tensor

\begin{theorem} From the whole  set of properties of the Riemann tensor, $R_{ijkl}+ R_{ijlk}=0$ and $R_{ijkl}+ R_{jikl}=0$ are separable. 
\end{theorem}

\noindent {\bf Proof:}   
We use the Riemann properties $P=\{(\ref{riemp01}),(\ref{riemp02}),(\ref{riemp1}), (\ref{riemp2}), (\ref{riemp3}), (\ref{riemp4})\}$. Let $P_1=\{p_1,p_2\}= \{(\ref{riemp01}),(\ref{riemp02})\}$ and $P_2=P-P_1$. 

Given any monomial having the Riemann tensor as a factor and using $P_1$ to improve the ordering, can be written as
\[
RF(i,j,k,l)=...R_{ijkl}...
\]
  where the indices $ijkl$ can be free or dummy but we can consider that  the actual names will correspond to the most orderly form. $P_1$ are separable {\it iff} $P_2(\{RF(i,j,k,l) \})= P_2(\{RF(i,j,l,k)\})= P_2(\{RF(j,i,k,l) \})$

Let us analyze $P_2$ case by case. We will use the notation ({\it ref.})$(\{M\})$ to indicate the set of properties $P_i(M)$ being $P_i$ a property of the set $P$:
 
\noindent{\bf pairsymmetrie (\ref{riemp1})}\newline
 Clearly $\{(\ref{riemp01}),(\ref{riemp02})\}$ is  separable from the set $\{(\ref{riemp01}),(\ref{riemp02}), (\ref{riemp1})\}$.

\noindent {\bf Bianchi I (\ref{riemp2})}\newline 
(\ref{riemp2})$(\{RF(i,j,k,l)\})$ gives
\begin{equation} 
RF(i,j,k,l)+ RF(i,l,j,k)+ RF(i,k,l,j)=0\label{b1}
\end{equation}

(\ref{riemp2})$(\{RF(i,j,l,k)\})$ gives $RF(i,j,l,k)+ RF(i,k,j,l)+RF(i,l,k,j)=0$ which after using $\{(\ref{riemp01}),(\ref{riemp02})\}$ coincides with (\ref{b1})\newline
(\ref{riemp2})$(\{RF(j,i,k,l)\})$ gives 
$RF(j,i,k,l)+ RF(j,k,l,i)+ RF(j,l,i,k)=0$ which after using $\{(\ref{riemp01}),(\ref{riemp02})\}$ gives  
\begin{equation}
-RF(i,j,k,l)- RF(j,k,i,l)+ RF(j,l,i,k)=0\label{riemp20}
\end{equation}
which does not coincide with (\ref{b1}) so $p_2$ is not separable from Bianchi I, but since we have also pairsymmetrie we can consider (\ref{riemp2})$(\{RF(k,l,i,j)\})$ which gives (\ref{riemp20}) so we can conclude that $\{(\ref{riemp01}),(\ref{riemp02})\}$ is separable from the set $\{(\ref{riemp01}),(\ref{riemp02}), (\ref{riemp1}),(\ref{riemp2})\}$.

\noindent {\bf Bianchi II (\ref{riemp3})}\newline
Considering a monomial of the form  
$RF(i,j,k,l,m)= .....R_{ijkl;m}....$ and finding (\ref{riemp3})$(\{RF(j,i,k,l,m)\})$ and (\ref{riemp3})$(\{RF(i,j,l,k,m)\})$ the properties generated are the same  as the property generated by (\ref{riemp3})$(\{RF(i,j,k,l,m)\}) $, so $\{(\ref{riemp01}),(\ref{riemp02})\}$ is separable from the set $\{(\ref{riemp01}),(\ref{riemp02}), (\ref{riemp1}),(\ref{riemp2}),(\ref{riemp3})\}$.

\noindent {\bf Ricci commuting relations (\ref{riemp4})}\newline
It is easy to observe that the (Riemann terms) of (\ref{riemp4}) have the same symmetry properties with respect to the indices $i,j,k,l$ as the term $R_{ijkl;m;n}$, specially $\{(\ref{riemp01}),(\ref{riemp02})\}$, so clearly  $\{(\ref{riemp01}),(\ref{riemp02})\}$ is separable from the set $\{(\ref{riemp01}),(\ref{riemp02}),(\ref{riemp1}),(\ref{riemp2}),(\ref{riemp3}),(\ref{riemp4})\}$, and this completes the proof.

{\it TTC} uses this separability property when internally defining the Riemann tensor.

In general we assume two sets of properties accordingly to the possibility of separating some from the whole set. We will call them $(n)$-properties  with $n=1,2$

\subsection*{How to handle $(n)$-properties }

Here we explain how {\it TTC} finds, solves and applies $(1)$-properties and $(2)$-properties. We will supose that we depart from a monomial $ M_1^{(0)}(I_F^{\beta_1})$ so let $n=1$ and go to {\bf step $n$}.  
\subsubsection*{Step $n$:}
Given a monomial $M_1^{(n-1)}(I_F^{\beta_1})$ the exhaustive application of the $(n)$-properties generates a set of relations of the type

\begin{eqnarray}
M^{(n-1)}_{i}(I_{F}^{\beta_p})+ \sum_{jq} b_{i j p q}M^{(n-1)}_{j}(I_{F}^{\beta_q})=0\ \ \ i,j=1,... \label{sistema2}
\end{eqnarray}
where $i,j,p,q$ enumerate possible $M,\beta,b$ objects appearing in (\ref{sistema2}) and $b_{...}$ are scalars.

This set of relations can always be solved in favor of the subset of the most orderly monomials 
\begin{equation}
M^{(n-1)}_{i}(I_{F}^{\beta_p})= \sum_{j,q} c_{i p j q}M^{(n)}_{j}(I_{F}^{\beta_{q}})
\end{equation}
$c_{i p j q}$ being  scalars and $M^{(n)}_{j}(I_{F}^{\beta_{q}})$ the most orderly monomials from the set $ M_{i}^{(n-1)}(I_{F}^{\beta_r})$ of (\ref{sistema2}) . 
The specific way to solve (\ref{sistema2})  will be explained in  section 4.

Finally we apply the obtained $(n)$-rules over the original monomial
\begin{equation}
M^{(n-1)}_1(I_{F}^{\beta_1})
\buildrel{(n)-{\rm rules} }\over{\Longrightarrow} 
\sum_{j,q} c_{1 1 j q}M^{(n)}_{j}(I_{F}^{\beta_{q}}) 
\end{equation}
$(n)$-rules are conveniently stored  for further use, as we will see later 

If $n<2$ set $n=n+1$ and go to {\bf step $n$}

\section{Ricci commuting relations}
As we have mentioned, Ricci commuting relations are inhomogeneous properties which must be handled with special care. If we want the algorithm to work correctly it is not sufficient to take into account Ricci commuting relation in the form (where we are taking an example of a single vector field $v$)  
\begin{equation}
v_{i;j;k}  \rightarrow v_{i;k;j}  + v_m \,R_{\,\,\,\,i\,j\,k}^m 
\end{equation}
since these rules will not benefit from other rules one might have $ v_{i;j...}$ in the case of monomials which do not have covariant derivatives explicitly, as in
\[
v_m \,R_{\,\,\,\,i\,j\,k}^m v_l 
\]

To solve this problem we have built a function, called \verb"AntiRicci" which takes any monomial and gives all the monomials related to it by Ricci commuting relations {\it via} covariant derivatives. After that we apply Ricci commuting relations over the covariant derivatives.
\begin{equation}
\verb"AntiRicci["v_m \,R_{\,\,\,\,i\,j\,k}^m\verb"]"=
\{ v_{i;j;k}\}
\end{equation}

With respect to separability we can write
\begin{theorem}
Given a tensor $T_{ijkl...}$ all symmetry properties of indices $(i,j,k,l..)$ are  separable from the Ricci commuting relations
\end{theorem}
\noindent {\bf proof:}
 Is trivial due to the fact that Ricci commuting relations are invariant under these symmetries.

 This Lemma does not apply for properties involving covariant derivative  indices. There is an important example when defining $v_i$ as Killing vector field: $v_{i;j}+ v_{j;i}=0$ is not separable from Ricci commuting relations.

The examples on the Appendix A are specially addressed these questions.  

Ricci commuting relations are internally included in the set of $(2)$-properties. Accordingly, if a tensor has no-separable properties and we want to use Ricci commuting relations properly, the set of properties must be included as $(2)$-properties

\section{Algorithm description }

The algorithm for simplification of polynomial tensors in index notation, called \verb+SimplifyAllIndex+, has the following functions:

\subsection*{Functions related to single tensors}
\begin{itemize}
\item \verb+InputTensor+:\\
\verb+ InputTensor[+{\it Tsymbol,basis,ranklist}\verb+]+ 
declares that the symbol ${\it Tsymbol}$ will be a tensor in the basis {\it basis} and with rank and type  {\it ranklist}. 

Example
 
\verb+InputTensor[T,XX,{1,1,1,1}]+

\item \verb+InputSymmetries[+{\it Tsymbol}\verb+[+{\it index}\verb+]+,{\it symmetriespecifications}\verb+]+
declares that the tensor, previously introduced using \verb+InputTensor+,
{\it Tsymbol} has the symmetry properties of the indices, positioned as indicated by the argument {\it index}, according to the specifications {\it symmetriespecifications}. 

Example 1:

\verb+InputSymmetries[T[i,j,k,l],{i,k,l}[1]]+ declares that  \verb+T+ is totally symmetric with respect to the indices  \verb+{i,k,l}+ and add this property to the set of $(1)$-properties

Example 2:

here we declare the same properties as the Riemann tensor, see (\ref{riemp01},\ref{riemp02},\ref{riemp1}, \ref{riemp2}, \ref{riemp3}, \ref{riemp4}), for the tensor \verb+T+ accordingly to the separability theorem given for the Riemann properties.  

\begin{verbatim}
InputSymmetries[T[i,j,k,l],
	{{i,j}}[1],{{k,l}}[1],{{i,j},{k,l}}[2],Cyclic[j,k,l][2]];

InputSymmetries[T[i,j,k,l,.;m],Cyclic[k,l,m][2]]
\end{verbatim} 

\item \verb+BasicRules[+{\it n}\verb+]+:
once the symmetries of indices have been declared the corresponding rules that implement these symmetries are stored in the list \verb+BasicRules[+{\it n}\verb+]+ . $ n=1$ for $(1)$-rules and  $n=2$ for $(2)$-rules. The value $n=0$ is reserved for the case of properties  that can be used directly , like $T_i^i=0$,  which are generally introduced explicitly by the user, or some properties {\it TTC} already has for Riemann tensor.
\end{itemize}

\subsection*{Functions related to monomials}
In the following we will assume that there are no direct properties stored in \verb+BasicRules[0]+ as they are inessential. 

The main function is \verb+SimplifyAllIndex+:

\verb+Index[+{\it metricname},\verb+SimplifyAllIndex[2]][+{\it polytens}\verb+]+ simplify, i.e. canonize, the polynomial {\it polytens} taking into account all the declared properties of the building tensor, as well as the dummy indices and the properties related to covariant derivatives.   
\verb+SimplifyAllIndex[2]+ is applied over each term of {\it polytens}. This function has, in turn, the following internal functions (We will assume that the initial monomial is $M(I_{F}^{\gamma},J_{D}^{\alpha})$):  

\begin{itemize}
\item \verb+SimplifyAllIndex[0]+: first we canonize the monomial with respect to the dummy indices. This step is taken each time a monomial is the result of some kind of index manipulation. So  our initial monomial and all other generated monomials will have the form $M_{i}^{(0)}(I_{F}^{\beta_p})$

\item \verb+LlistaMonomis[+{\it label,n,counter(n) }\verb+]+: 
in these lists all the monomials considered before the present case are stored in {\it globacodification} version together with its $\beta$-permutation which relate it to the way they are present in the corresponding rules. 
$n=1$ for $(1)$-rules and $n=2$ for $(2)$-rules. {\it counter(n)} is a counter which says that the monomials therein are the {\it counter(n)}-th studied monomials.

\item \verb+LlistaRules[+{\it label,n,counter(n)}\verb+]+: 
in these lists the simplifying rules of the monomial in the corresponding \verb+LlistaMonomis+ are stored. 
\end{itemize}

The specific way {\it TTC} stores $(n)$-rules is \newline

\noindent \verb"LlistaMonomis["{\it label,n,counter(n)}\verb"]=" 
$\left\{..., \left\{ M_i^{(n-1)} (I_F),\beta_i  \right\},... \right\} 
$\newline

\noindent \verb"LlistaRules["{\it label,n, counter(n)}\verb"]=" 
$
\left\{...,   
\left\{ M_i^{(n-1)} (I_F^{\beta_p}) \to 
\sum_{(j,q)} b_{i p j q} M_j^{(n)} (I_F^{\beta_q}) \right\},...
\right\}  
$

\begin{itemize}
\item \verb+SimplifyAllIndex[{+{\it n}\verb+}]+, with $n=1,2$,
simplify a monomial already simplified with respect to $n-1$ properties, say $M^{(n-1)}(I_F^{\gamma})$, with respect to $(n)$-properties. 
It searches if its {\it globalcodification} version, $M^{(n-1)}(I_F)$, is in  one of the corresponding lists \verb"LlistaMonomis" and if it is the case it takes, if necessary, the corresponding $(\gamma^{-1}\circ\beta_k)$ permutation version of the \verb"LlistaRules" solving with respect to the ordering criterion and applying the new rules to the monomial $M^{(n-1)}(I_F^{\gamma})$.

If $M^{(n-1)}(I_F)$ is not found in any lists \verb"LlistaMonomis" it applies the rules stored in \verb+BasicRules[+{\it n}\verb+]+ exhaustively  generating new lists \verb"LlistaMonomis" and  \verb"LlistaRules" and applying them as explained above.

\item \verb+PFind+ is a simple tool that solves any system of linear equations with a fixed  ordering criterion.

Let $\{x_{1},x_{2},...,x_{M}\}$ be the space of ordered variables to be solved. Let 
\begin{equation}
\verb+eqlist+=\{c^{1}+\sum_j a^{1 j} x_{j}=0,c^{2}+\sum_j a^{2 j} x_{j}=0,...,c^{N}+\sum_j a^{N j} x_{j}=0 \}  
\end{equation}
be the list of $N$ equations with $M$ variables ordered first taking into account the variables and then the coefficients. That is to say, if in the $n$-th equation the most disordered variable is  $x_{m}$ then $x_{m+1}$ cannot be present in the $(n-1)$-th equation. 

\verb+PFind+ takes the initial list of equations and starts the following loop:

{\bf Step $1$:}
\verb+PFind+ takes the first equation in \verb+eqlist+. It is solved with respect to the most disordered variable. This result is stored in \verb+eqrule[1]+.
\verb+eqrule[1]+ is applied over \verb+eqlist+ and after eliminating zeros this result is assigned to \verb+eqlist+. Go to step 2

{\bf Step $n>1$:}
If \verb+eqlist+ is empty the loop is finished. On the contrary \verb+PFind+ takes the first equation in \verb+eqlist+. It is solved with respect to the most disordered variable. This result is stored in \verb+eqrule[+{\it n}\verb+]+.
\verb+eqrule[+{\it n}\verb+]+ is applied over \verb+eqrule[+{\it i}\verb+]+ $i=1,...,n-1$. \verb+eqrule[+{\it n}\verb+]+ is applied over \verb+eqlist+, the zeros are eliminated and the result assigned to \verb+eqlist+. Go to the step $n+1$.

\verb+eqlist+ will remain empty if the system is compatible. These must be the case for a set of compatible tensor properties. 
\end{itemize}

The computer time and memory of \verb"PFind" have been controlled. 

\section{Concluding Remarks}

We have solved the problem of simplification of tensor polynomials in index notation with respect to a large set of properties for the building tensor satisfactorily. The computer time and memory needed are sufficiently reasonable so we think that the program can be of interest for practical purposes. We have successfully used the algorithms presented \cite{XACollinson} 

Due to the generic character of the algorithm (the only requirement is that the input polynomial follows the usual mathematical rules) it can be useful in the kind of fields where tensors are a natural language. As is well known tensor calculus is powerful enough to cover large fields of mathematics, physics and engineering.

There is an important  set of properties not included in the present algorithm. These are the ones that appear when the dimension of the space is taken into account. 
The essential reason for not including these relations is that applying properties of the building tensor do not generate this kind of relations. They arise when the whole monomial is taken into account so we need some tool to generate the dimensional properties for a given monomial. We are leaving this new tool for future work \cite{XADim}.

\section*{Acknowledgments}
X.J. would like to thank the Comisi\' on Asesora de Investigaci\' on Cient\' \i fica y T\' ecnica for partial financial support, under Contract No. PB96-0384. The authors would also like to thank the Laboratori de F\' \i sica Matem\` atica at the Societat Catalana de F\' \i sica for partial financial support.

\section*{Appendix A}

We present here an example of how the algorithm \verb+SimplifyAllIndex+ can be applied. 
The session is part of a {\it Mathematica} notebook. The inputs are indicated by 
\verb+In[]:=+. The outputs have no prefix. The titles and comments are enclosed using the symbols \verb+(*+{\it comment}\verb+*)+. 

Since we use the Riemann tensor as an example we introduce this tensor using the {\it TTC} function \verb+InputSRiemann+. This is internally equivalent to use \verb+InputTensor+ and \verb+InputSymmetries+ to declare the Riemann and Ricci tensors and to give the relation between them and between the curvature scalar. All tensor calculus conventions follow (\cite{Misner}). 

In this first short academic example we first introduce coordinate system \verb"cx4" and a metric \verb"gn" together with the Riemann tensor.

We define the tensor $V$ and $L$ with symmetries $L^{ijk}= L^{kji}$ , $L^{ijk;m}= L^{jik;m}$ and $L^{ijk;m}+L^{imj;k}+L^{ikm;j}=0$. Note that $L$ have not separable properties so we will define all properties as a (2)-properties.

Next we declare the indices for outputs and define and simplify  a monomial

{\fontsize{8}{8}
\begin{verbatim}

In[]:=<<ttc.m
 ----------------------------------------
 |  Tools  of  Tensor  Calculus 4.1.0   |
 |  A.Balfagon,P.Castellvi and X.Jaen   |
 |     http://baldufa.upc.es/ttc        |
 |     e-mail:ttc@baldufa.upc.es        |
 |    version: september, 29,1999       |
 ----------------------------------------
 |     Session started on               |
 |     October,   5 , 1999              |
 |     at 11 h 15 min 48 s              |
 ----------------------------------------

In[]:=(
InputCoordinates[cx4,4];
InputSMetric[gn,cx4,"g",g];
InputSRiemann[gn,cx4,"R",Rie,Ric,R];
InputTensor[{L},cx4,{1,1,1}];
InputTensor[{V},cx4,{1}];
InputSymmetries[L[i,j,k],{i,k}[2]];
InputSymmetries[L[i,j,k,.;m],{i,j}[2],
	Cyclic[j,k,m][2]];
InputIndex[{a,b,c,d,e,m,n}]
);

3/2 L[i,j,k] Rie[-i,-j,q,p] V[-k] //Index[gn]

           c d e             a b 
        3 L       V     R
 a b               e      c d 
0     + ---------------------------
                     2
%//Index[gn,SimplifyAllIndex[2]]

           b       d    c a 
        3 L       V   R
 a b         c d     
0     - ----------------------- + 
                   4
 
       b     d      c a 
  3 L       V      R
     c   d    
  ----------------------- + 
             4
      a       d    c b 
  3 L       V    R
       c d    
  ----------------------- - 
             4
 
       a     d      c b 
  3 L       V      R
     c   d     
  ----------------------- + 
             4
 
       a b        c d 
  3 L       V    R
     c       d 
  ----------------------- - 
             4
 
       b a        c d 
  3 L       V    R
     c       d 
  -----------------------
             4
 
(* Note that if Ricci=0 then the original monomial
 is zero, which is not obvious from the beginning*)
\end{verbatim}

Next we show how the same kind of calculation is able to find a known result: The vanishing of the magnetic part of the Weyl tensor for static space-times, see \cite{Kramer}
\begin{verbatim}
(* first we introduce the predefined volume form which we name  H*)
In[]:=
InputSAForm[gn,cx4,"H",H];

(*Here we introduce the Killing field Vt, which we define as
a vector antisymmetric between his own index and
the covariant derivative index. *)

In[]:=
(InputTensor[Vt,cx4,{1}];
InputSymmetries[Vt[a,.;b],{{a,b}}[2]]);

(*Next we introduce the static condition. Since this
property involves more than one tensor, it must be introduced 
explicitly by the user and taking into account 
its differential consequences.
We introduce these conditions in BasicRules[2] 
using the specific syntaxis a_:>(a;b) instead of a_:>b !!*)
 
(BasicRules[2]=Join[BasicRules[2],{

Vt[a_] Vt[b_,.;c_]:>(Vt[a] Vt[b,.;c];-
           Vt[c] Vt[a,.;b]-Vt[b] Vt[c,.;a]),

Vt[a_] Vt[b_,.;c_,.;d_]:>(Vt[a] Vt[b,.;c,.;d];
        -Vt[c] Vt[a,.;b,.;d]-Vt[b] Vt[c,.;a,.;d]-
              Vt[a,.;d] Vt[b,.;c]-Vt[b,.;d] Vt[c,.;a]-Vt[c,.;d] Vt[a,.;b]),


Vt[b_,.;d_] Vt[c_,.;a_]:>(Vt[b,.;d] Vt[c,.;a];
       -Vt[c] Vt[a,.;b,.;d]-Vt[b] Vt[c,.;a,.;d]-
                 Vt[a] Vt[b,.;c,.;d]-Vt[a,.;d] Vt[b,.;c]-Vt[c,.;d] Vt[a,.;b]),

}]);

(* Next we define Weyl Tensor*)
In[]:=
IndexUpdate[Weyl,":=",
0[a,b,c,d]+Rie[a,b,c,d]-1/2 g[a,c] Ric[b,d]+
1/2 g[a,d] Ric[b,c]+1/2 g[b,c] Ric[a,d]-1/2 g[b,d] Ric[a,c]+
R/12(g[a,c] g[b,d]-g[b,c]g[a,d]+g[b,d]g[a,c]-g[a,d]g[b,c])
//Index[gn]]

(*Next we define the magnetic part of Weyl, BW*)
In[]:=
IndexUpdate[BW,":=",
1/2 H[a,r,c,m] Weyl[-c,-m,b,n] Vt[-r] Vt[-n]//Index[gn]];  

(*We can see BW expanded. Now we like to use {i,j,k,l...} for outputs*)
In[]:=InputIndex[{i,j,k,l,m,n,o,p}];

In[]:=BW[-a,-b]//Index[gn,SuperIndexExpand]
 
                      l   k                   l k                   l k m 
        R Vt   Vt   H           R Vt   Vt   H           Vt   Vt   H         R
            k    l   i   j          k    l   i     j      k    l   i         m j 
0     + --------------------- - --------------------- + ------------------------- - 
 i j             12                      12                         4
 
               l m k                       l   m     k                 l m       k 
  Vt   Vt   H         R       Vt   Vt   H         R       Vt   Vt   H         R
    k    l   i         m j      k    l   i   j     m        k    l   i     j   m 
  ------------------------- - ------------------------- + ------------------------- + 
              4                           4                           4
 
               l m n         k 
  Vt   Vt   H         R
    k    l   i         m n j 
  -----------------------------
              2

(*Finally we simplify BW, which give the final result*) 

In[]:=%//Index[gn,SimplifyAllIndex[2]]

0

(*Here we can see the internal aspect of some of the stored rules*)

In[]:=LlistaMonomis[gn,cx4,2,2,2]


    IndexD[1]    IndexF[1] .;IndexD[2] .; IndexD[3]
{{Vt           Vt
 
    IndexD[1] IndexD[2] IndexD[3] IndexF[2]
  H                                           , {2, 1}}, 
 
     IndexD[1]    IndexD[2]  IndexD[1] IndexD[3] IndexD[4] IndexF[1]
  {Vt           Vt          H 
   
  IndexD[2] IndexF[2] IndexD[3] IndexD[4] 
 R                                         , {1, 2}},.....


In[]:=
LlistaRules[gn,cx4,2,2,2]


    IndexD[1]    IndexF[2]   IndexD[2] IndexD[3] IndexD[4] IndexF[1] 
{Vt           Vt            H
 
      IndexD[1] IndexD[2] IndexD[3] IndexD[4]
    R                                         -> 0, .....
 
\end{verbatim}

\end{document}